\documentclass[epj,twocolumn]{webofc}
\usepackage[varg]{txfonts}
\usepackage{amsmath}
\usepackage{graphicx}
\usepackage{float}
\usepackage{hyperref}
\begin{document}

\title{NESSA: a compact 14~MeV D--T neutron source facility at Uppsala University}

\author{
  \firstname{Sandipan} \lastname{Dawn} \inst{1}
  \and
  \firstname{Elias} \lastname{Arnqvist} \inst{1}
  \and
  \firstname{Göran} \lastname{Ericsson} \inst{1}
  \and
  \firstname{Linus} \lastname{Hägg} \inst{1}
  \and
  \firstname{Stefan} \lastname{Jarl-Holm} \inst{1}
  \and
  \firstname{Mattias} \lastname{Lantz} \inst{1}
  \and
   \firstname{Andreas} \lastname{Lindner} \inst{2}
  \and
  \firstname{Erik} \lastname{Andersson Sundén} \inst{1}
  \and
  \firstname{Stephan} \lastname{Pomp}\inst{1}\thanks{\email{stephan.pomp@physics.uu.se}}
}

\institute{
  Division of Applied Nuclear Physics, Department of Physics and Astronomy, Uppsala University, 751\,20 Uppsala, Sweden
\and
  Department of Electrical Engineering, Uppsala University, 751\,03 Uppsala, Sweden
}

\abstract{%
The NESSA (Neutron Source in Uppsala) facility hosts a compact 14\,MeV deuterium--tritium sealed tube neutron generator at the \AA ngstr\"om Laboratory, Uppsala University. The generator, housed in a bunker inside the FREIA hall, reaches a maximum yield of $4.7\times10^{8}$\,n/s. This paper describes the facility: the generator, the bunker and its shielding, the detector systems, and the Monte Carlo models used to characterize the neutron field. We also report the first commissioning measurements: yield calibration with $^{93}$Nb activation foils, fission chamber response at two positions, and simulated air and structural activation. Initial indium foil activations and single event effect (SEE) tests on silicon devices are also presented. The facility will be used for nuclear data measurements, neutron detector response studies, moderation and thermalization experiments, irradiation testing of electronics as well as for training and education.
}
\maketitle
\section{Introduction}
\label{sec:intro}
Several fast neutron facilities in Sweden and the Nordic region have closed in recent years, including The Svedberg Laboratory (TSL) in Uppsala, the Studsvik reactor, and the Halden reactor in Norway. NESSA (Neutron Source in Uppsala) is built at the Division of Applied Nuclear Physics, Uppsala University, to restore regional access to a high yield fast neutron source. The facility sits in the FREIA (Facility for Research Instrumentation and Accelerator Development) hall of the \AA ngstr\"om Laboratory, a local infrastructure, equipped for testing accelerator components, RF cavities, magnets, etc., and already licensed for radiation work.

\begin{figure}[h]
\centering
\includegraphics[width=0.85\columnwidth]{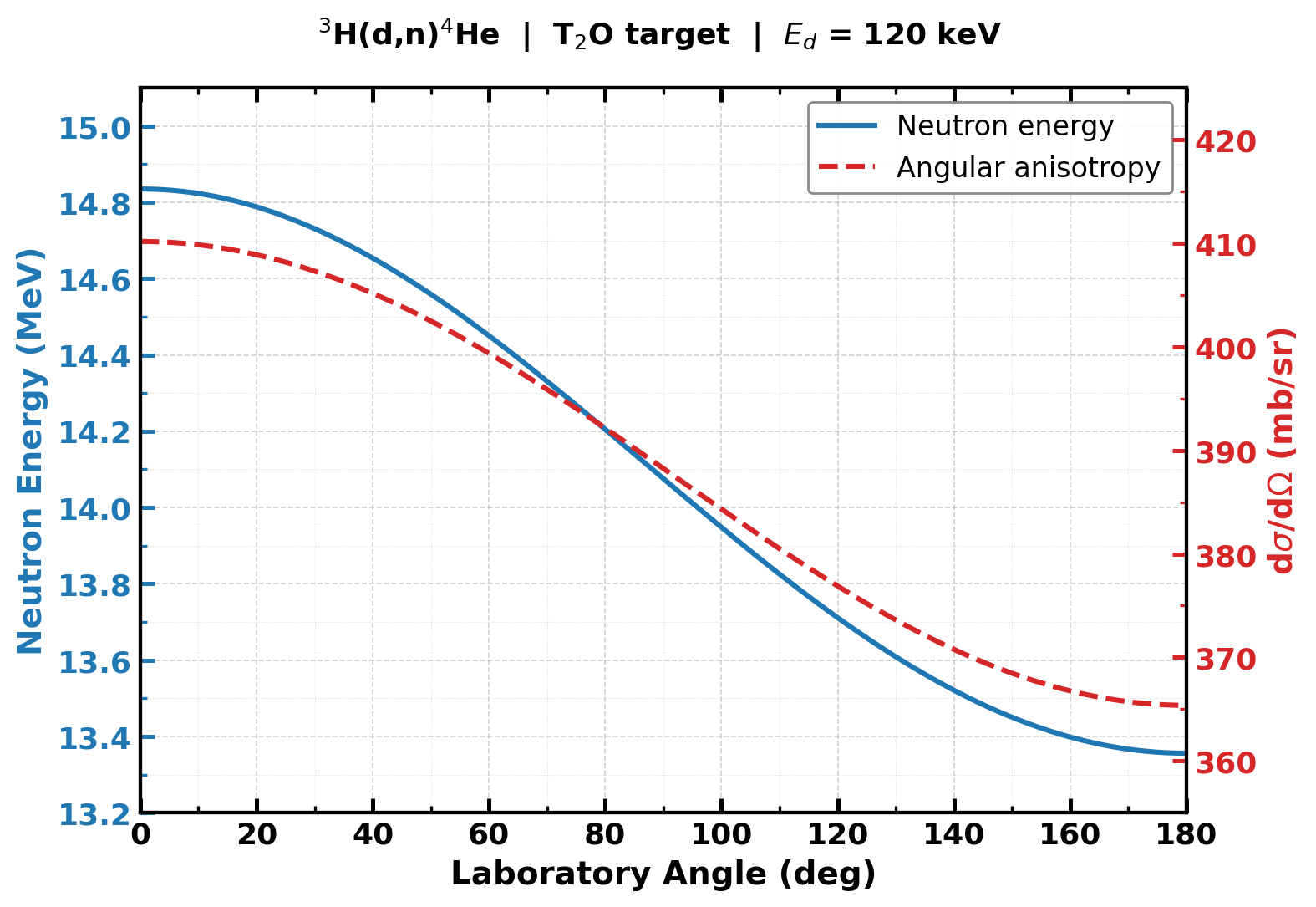}%
\caption{Neutron energy and angular anisotropy at different angles in the laboratory frame.}
\label{fig:anisotropy}
\end{figure}

NESSA is built around a D-T generator, which produces quasi-monoenergetic 14.1\,MeV neutrons from the T(d,n)$^{4}$He reaction, with a cross section of about 5\,barn at 100\,keV deuteron energy in the center of mass frame~\cite{leung2004}. The current generator is a Sodern GENIE\,16, producing up to $4.7\times10^{8}$\,n/s into $4\pi$. The facility is designed to eventually host a higher-yield generator ($\sim10^{10}$\,n/s) with maximum fluxes up to $10^{9}$\,n\,cm$^{-2}$\,s$^{-1}$ at the closest user position (CUP). The generator can be operated in both DC and pulsed modes (minimum pulse width 10\,$\mu$s), and the neutron yield can be tuned from $1\times10^{7}$\,n/s to maximum. At different angular positions and operating voltages, reaction kinematics produce an anisotropy in both energy and flux; the angular dependence is studied with DROSG-2000~\cite{drosg2000} and shown in Fig.~\ref{fig:anisotropy}.

\section{Facility and instrumentation}
\label{sec:facility}
\begin{figure}[t]
\centering
\includegraphics[height=0.60\columnwidth]{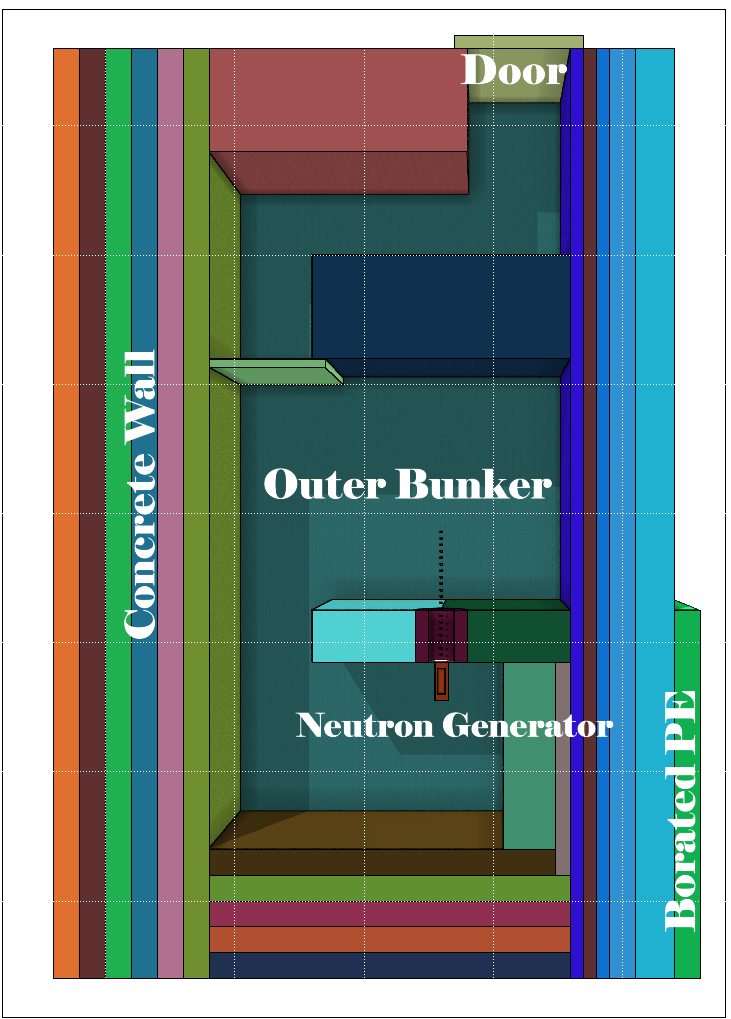}%
\hfill%
\includegraphics[height=0.57\columnwidth, scale=0.82]{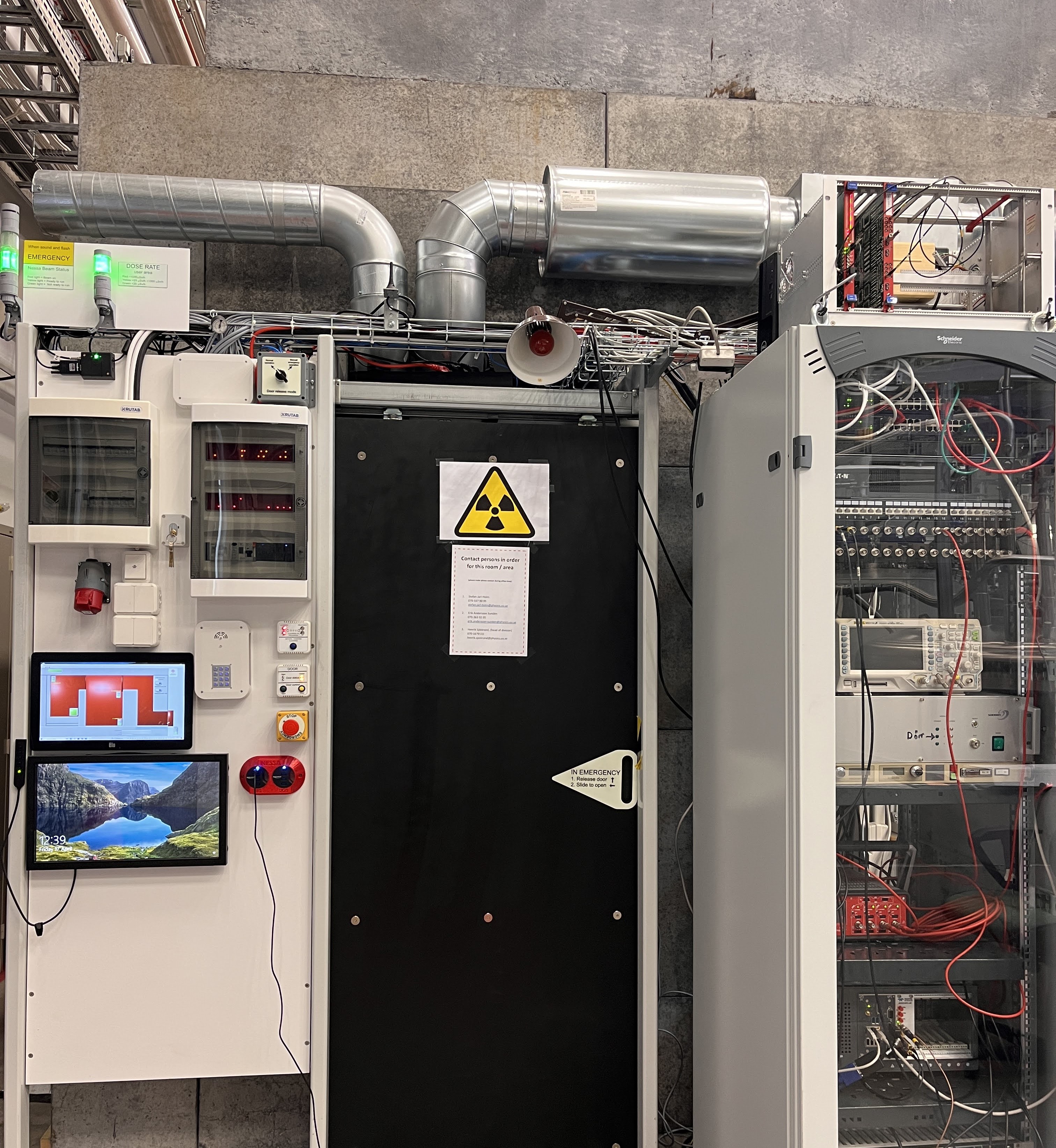}
\caption{FLUKA reconstructed model of the NESSA irradiation facility showing the outer bunker, neutron generator, concrete, iron and lead walls, borated polyethylene layers and access door (left). Photograph of the entrance to the NESSA experimental area (right).}
\label{fig:geometry}
\end{figure}

The NESSA irradiation facility consists of an inner bunker that houses the generator and the close user irradiation positions, along with a larger outer bunker for lower flux experiments (Fig.~\ref{fig:geometry}). Reinforced concrete walls, iron and lead linings at specific locations, form the primary biological shielding. Polyethylene and borated polyethylene panels are added to selected inner surfaces: the hydrogenous material moderates fast neutrons, and the boron absorbs the resulting thermal neutrons with little secondary gamma emission. Access is controlled by various interlock systems. Neutron dose rate meters, scintillators, and GM-based gamma counters are installed inside the bunker to monitor dose rates in real time. Earlier neutron transport studies using Total Monte Carlo methodologies~\cite{sjostrand2014} served as a reference for the shielding layout.

Several detector systems are installed at NESSA for various experimental requirements. Two Centronic FC4A micro fission chambers ($\mu$FC), one with a $^{235}$U deposit and one with $^{238}$U (4\,mg U$_{3}$O$_{8}$ each), are used for real-time flux monitoring. They are read out with Mesytec MPR-1 preamplifiers and digitized with a Teledyne SP Devices ADQ14 card (1\,GSPS, 14-bit). NaI(Tl) and LaBr$_{3}$ scintillators are available for prompt gamma spectroscopy, and a pair of EJ-301 liquid scintillators is used for fast-neutron detection with n/$\gamma$ pulse-shape discrimination. Two co-axial HPGe gamma spectrometers, nUGGLA (n-type) and UGGLA~\cite{elias2024} (p-type), are used for spectrometry of neutron activation experiments.

\section{Neutron yield and field characterization}
\label{sec:characterisation}

The absolute yield of the generator was determined by the activation of a niobium foil via the $^{93}$Nb(n,2n)$^{92m}$Nb reaction ($T_{1/2}=10.15$\,d, $E_{\gamma}=934.44$\,keV, $I_{\gamma}=99.15\%$)~\cite{nndc}. A foil of 0.345\,g, 10\,mm diameter, and 0.5\,mm thickness was irradiated for 4\,h at $r=4.75$\,cm and $90^{\circ}$ to the beam direction. The end-of-irradiation activity is related to the neutron flux at the foil position by
\begin{equation}
A_{0} \;=\; N\,\sigma\,\Phi\,S\,C_{\mathrm{ss}},
\label{eq:activation}
\end{equation}
where $N$ is the number of target atoms in the foil, $\sigma$ is the reaction cross section at the neutron energy of interest, $\Phi$ is the flux at the foil, $S=1-e^{-\lambda t_{\mathrm{irr}}}$ is the saturation factor that accounts for the build-up and simultaneous decay of the product nucleus during irradiation time $t_{\mathrm{irr}}$ (with $\lambda$ as its decay constant), and $C_{\mathrm{ss}}$ is the self-shielding correction for the attenuation of incoming neutrons within the foil. The source yield is then obtained from $\Phi$ after correcting for the finite foil size through a geometry factor $C_{\mathrm{geom}}$ and for the angular anisotropy of the D-T reaction. With $\sigma=0.462$\,b from ENDF/B-VIII.0~\cite{endf}, $S=0.0114$, and a combined self-shielding correction $C_{\mathrm{total}}=1.021$, the flux at the foil is $\Phi=(1.51\pm0.10)\times10^{6}$\,n\,cm$^{-2}$\,s$^{-1}$. After applying $C_{\mathrm{geom}}=1.011$ and the source anisotropy, the yield is $Y=(4.33\pm0.29)\times10^{8}$\,n/s. This agrees well with the factory value of $4.4\times10^{8}$\,n/s, keeping in mind that the tube has been in storage for several years and some output degradation is expected.

The neutron field inside and around the irradiation facility has been modeled in four independent Monte Carlo codes — FLUKA~\cite{fluka2022}, Geant4~\cite{agostinelli2003geant4} (v11.4, QGSP\_BIC physics list), MCNP~\cite{shultis2013mcnp}, and PHITS~\cite{niita2006phits}. Running four codes in parallel allows cross-checking of physics models, nuclear data libraries, and geometry implementations — particularly valuable for a facility that will host benchmark experiments. Predicted fluence spectra and integrated fluxes at the CUP agree between codes within a few percent at the 14\,MeV peak and within about 20\% in the scattered low-energy tail.

\begin{figure}[t]
\centering
\includegraphics[width=0.48\columnwidth]{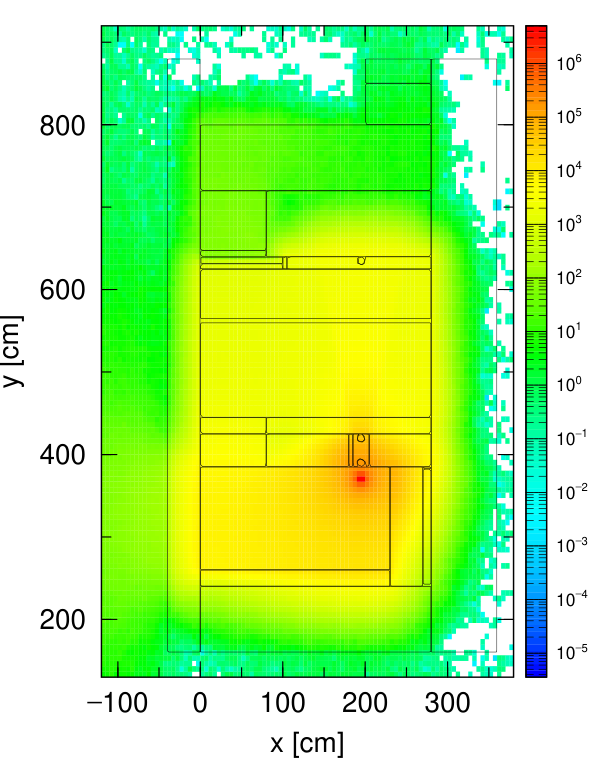}%
\hfill%
\includegraphics[width=0.48\columnwidth]{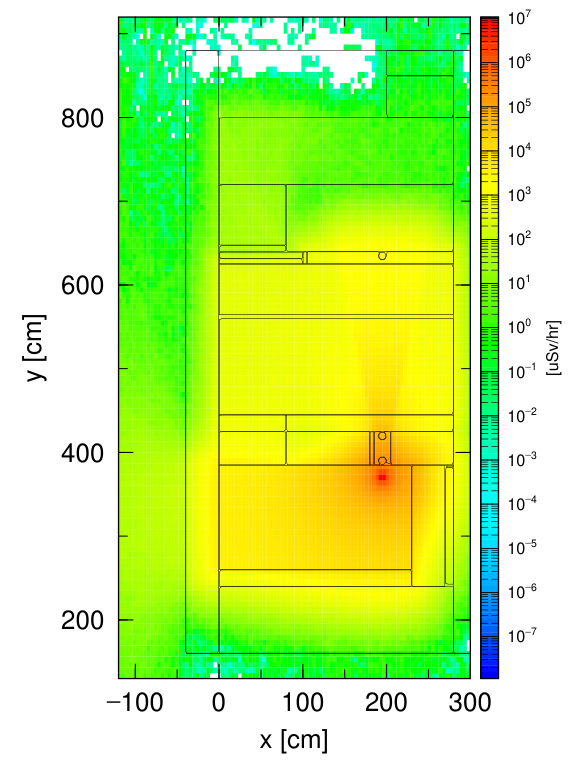}
\caption{(a) Simulated neutron fluence (n\,cm$^{-2}$\,s$^{-1}$) and (b) ambient neutron dose equivalent rate H*(10), both scaled to $4.7\times10^{8}$\,n/s, in the horizontal plane through the source.}
\label{fig:fluence_dose}
\end{figure}

Figure~\ref{fig:fluence_dose} shows the 2D PHITS simulated neutron fluence and ambient dose-equivalent rate H*(10) maps scaled to $4.7\times10^{8}$\,n/s. At the CUP the fluence reaches about $3\times10^{6}$\,n\,cm$^{-2}$\,s$^{-1}$ at the present yield, and would reach $\sim10^{9}$\,n\,cm$^{-2}$\,s$^{-1}$ ($\sim10^{12}$\,n\,cm$^{-2}$\,h$^{-1}$) with the planned upgrade, sufficient for meaningful single event effect testing. The outer bunker sits at $10^{4}$--$10^{5}$\,n\,cm$^{-2}$\,s$^{-1}$, suitable for activation and dosimetry work. The neutron dose rate close to the source reaches $\sim$1\,Sv\,h$^{-1}$ at maximum yield; outside the outer concrete wall the predicted dose stays below 1\,$\mu$Sv\,h$^{-1}$, within the public exposure limit, due to the combined effect of moderation in polyethylene, capture in borated polyethylene, and attenuation in primary shielding.

\begin{figure}[t]
\centering
\includegraphics[width=\columnwidth]{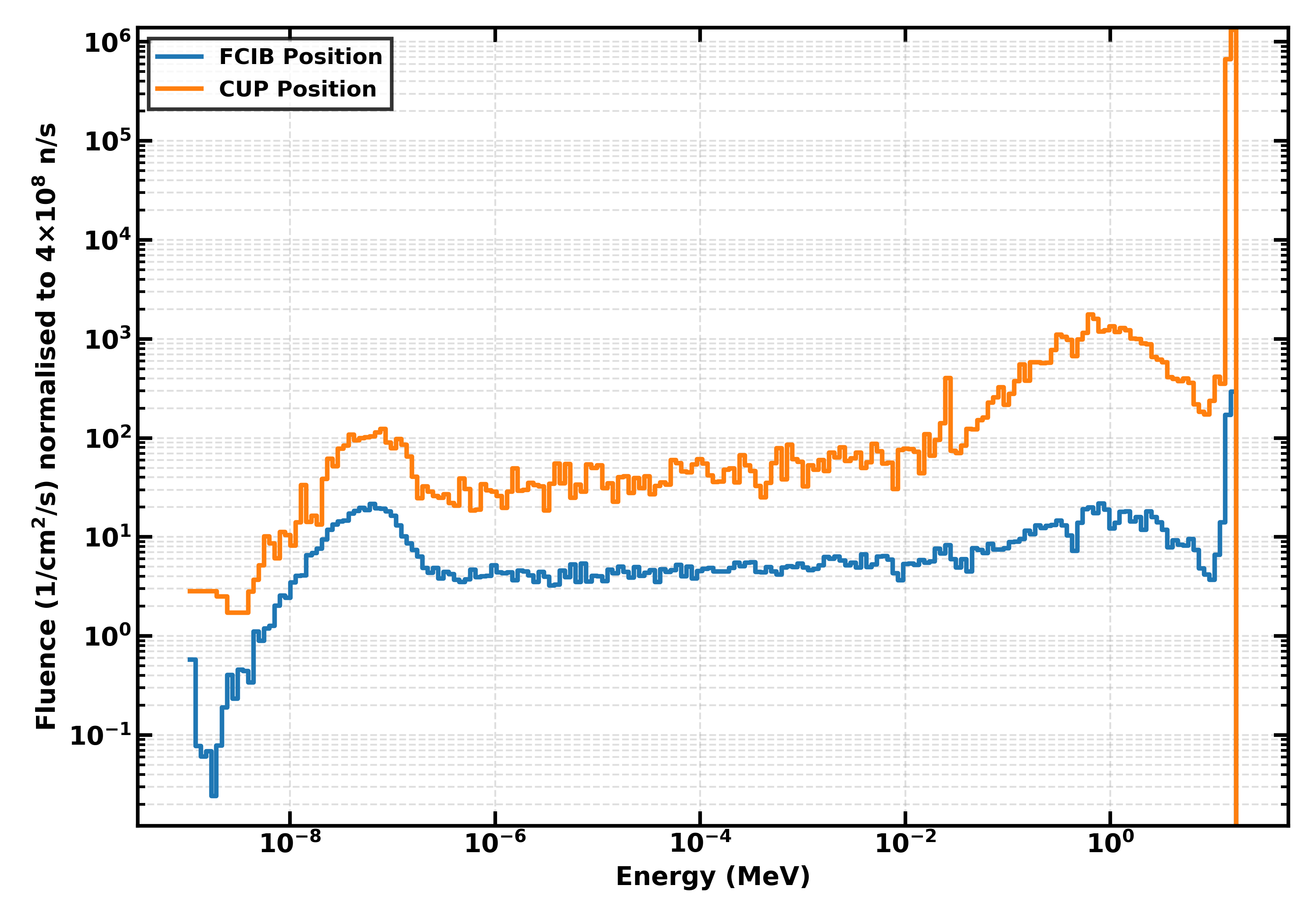}
\caption{PHITS simulated neutron spectra at the CUP (orange) and the fission-chamber inner bunker position FCIB (blue), normalized to $4.7\times10^{8}$\,n/s.}
\label{fig:spectra}
\end{figure}

The PHITS simulated spectra at two representative positions are shown in Fig.~\ref{fig:spectra}. At the CUP the 14\,MeV peak dominates. At the fission chamber inner bunker position (FCIB, 130\,cm from the source near the wall), the 14\,MeV component is about four orders of magnitude lower, while the room-scattered component from thermal to MeV energies drops by only about one order. This matters for detector response studies: the $^{235}$U fission chamber at FCIB is driven predominantly by low-energy scattered neutrons, since $\sigma_{f}(^{235}\mathrm{U})$ is weighted toward low energies.

The $\mu$FC pair was measured at CUP (40\,mm, 4\,h run, $4.4\times10^{8}$\,n/s) and at FCIB (130\,cm, 20\,h run, $1.8\times10^{8}$\,n/s)~\cite{ctreport}. Pulse-height spectra were corrected for background and losses below the discriminator threshold. Predicted rates were obtained from $C_{R}=\varepsilon\,\sigma_{\mathrm{avg}}\,N\cdot n\,v$ using simulated spectra, JENDL-5 cross sections, and detector efficiency $\varepsilon=0.55$. For $^{235}$U the predicted rates are 31.7 and 2.02\,cps at CUP and FCIB, against measurements of 34.2 and 2.21\,cps — agreement at the 10\% level. The $^{238}$U deviations are larger (22--52\%), driven by uncertainties in the sub-threshold correction. The spectral shift between the two positions is well reproduced: $\sigma_{\mathrm{avg}}$ for $^{235}$U changes from 2.55\,b at CUP to 92.3\,b at FCIB, consistent with the simulated spectra.

\section{Activation and radiological safety}
\label{sec:safety}

Continuous generator operation activates the structural materials of the bunker (steel, aluminum, concrete, stainless steel hardware) and the air in the room. Monte Carlo simulations incorporating radioactive decay identify the main products. The short-lived group : $^{28}$Al (2.24\,min), $^{52}$V (3.74\,min), $^{27}$Mg (9.46\,min), $^{56}$Mn (2.58\,h), $^{24}$Na (14.96\,h) - governs immediate access after long irradiation. After 24\,h at maximum yield, the photon dose rates from $^{28}$Al and $^{24}$Na immediately after shutdown are 6.30 and 5.54\,$\mu$Sv/h, respectively. Medium-lived isotopes $^{59}$Fe (44.5\,d), $^{58}$Co (70.9\,d), and $^{54}$Mn (312\,d) set the residual dose over weeks to months; long-lived $^{60}$Co (5.27\,y) and $^{22}$Na (2.60\,y) appear at trace levels but are tracked for end-of-life classification. For a typical 4\,h run at maximum yield, the contact dose at the generator surface drops from 47\,$\mu$Sv/h at 0.15\,hr after shutdown to 0.5\,$\mu$Sv/h after one hour, defining the practical cool-down window for in-bunker maintenance.

Air activation was projected for a 24\,h run at $10^{10}$\,n/s, representative of the planned higher-yield generator (Table~\ref{tab:airact}). The dominant product at beam-off is $^{13}$N from $^{14}$N(n,2n), which decays quickly and drops below $10^{2}$\,Bq within an hour. $^{41}$Ar from $^{40}$Ar(n,$\gamma$) is the persistent component; its slow decay is managed by the forced ventilation system, hard-wired into the generator safety chain and running automatically whenever the beam is on.

\begin{table}[t]
\centering
\caption{Air activation products (Bq, $\pm1\sigma$) at 1\,min and 30\,min after beam shutdown, following 24\,h irradiation at $10^{10}$\,n/s. Isotopes below 100\,Bq are omitted.}
\label{tab:airact}
\begin{tabular}{lccc}
\hline
\textbf{Isotope} & \textbf{$T_{1/2}$} & \textbf{1\,min} & \textbf{30\,min} \\
\hline
$^{13}$N  & 9.97\,min  & $(6.4\pm0.1){\times}10^{3}$ & $(8.6\pm0.2){\times}10^{2}$ \\
$^{41}$Ar & 109.6\,min & $(1.3\pm0.07){\times}10^{3}$ & $(1.1\pm0.06){\times}10^{3}$ \\
$^{16}$N  & 7.13\,s    & $(217\pm0.6)$ & $<1$ \\
\hline
\end{tabular}
\end{table}

\section{Initial experiments and analysis}
\label{sec:initial}

To demonstrate the types of experiments NESSA can support, we carried out a series of measurements covering nuclear data, gamma-ray spectroscopy, and electronics testing. Monte Carlo simulations supported each stage of the analysis: corrections for foil self-shielding, detector response and efficiency, and separation of reaction products with overlapping gamma lines.

We measured the cross section for $^{115}$In(n,2n)$^{114m}$In at 14\,MeV using natural indium foils. The result, $1210\pm81$\,mb, is consistent with recent literature~\cite{sunitha2020}. A more challenging measurement is the isomeric yield ratio $^{112m}$In/$^{112g}$In from the $^{113}$In(n,2n)$^{112}$In reaction on the 4.3\% minor isotope of natural indium. The two states differ in spin: ground state $J^{\pi}=1^{+}$ and isomer at 156.61\,keV with $J^{\pi}=4^{+}$ and $T_{1/2}=20.67$\,min . This isomeric ratio reflects how the compound nucleus distributes its angular momentum during gamma decay — a sensitive probe of nuclear level density. The difficulty lies in spectral overlap at 617.5\,keV: both $^{112g}$In ($T_{1/2}=14.97$\,min) and $^{112}$Ag ($T_{1/2}=3.13$\,h) from the $^{115}$In(n,$\alpha$) channel emit at this line. We acquired HPGe spectra at multiple times after irradiation. At late times, the short-lived $^{112g}$In has decayed, leaving only $^{112}$Ag at 617.5\,keV. We used this to determine the $^{112}$Ag contribution at earlier times, subtracted it from the total 617\,keV yield, and isolated the $^{112g}$In component. The measured isomeric ratio $3.87\pm0.14$ agrees with published values~\cite{luo2017}.

\begin{figure}[t]
\centering
\includegraphics[width=0.7\columnwidth]{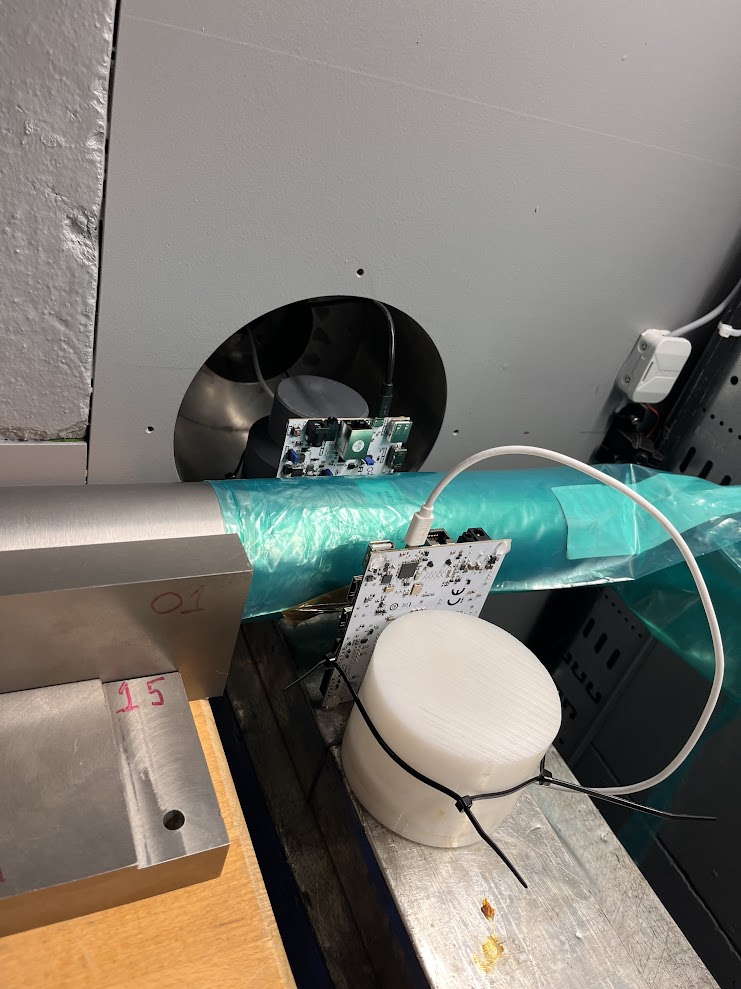}
\hfill%
\includegraphics[width=0.9\columnwidth]{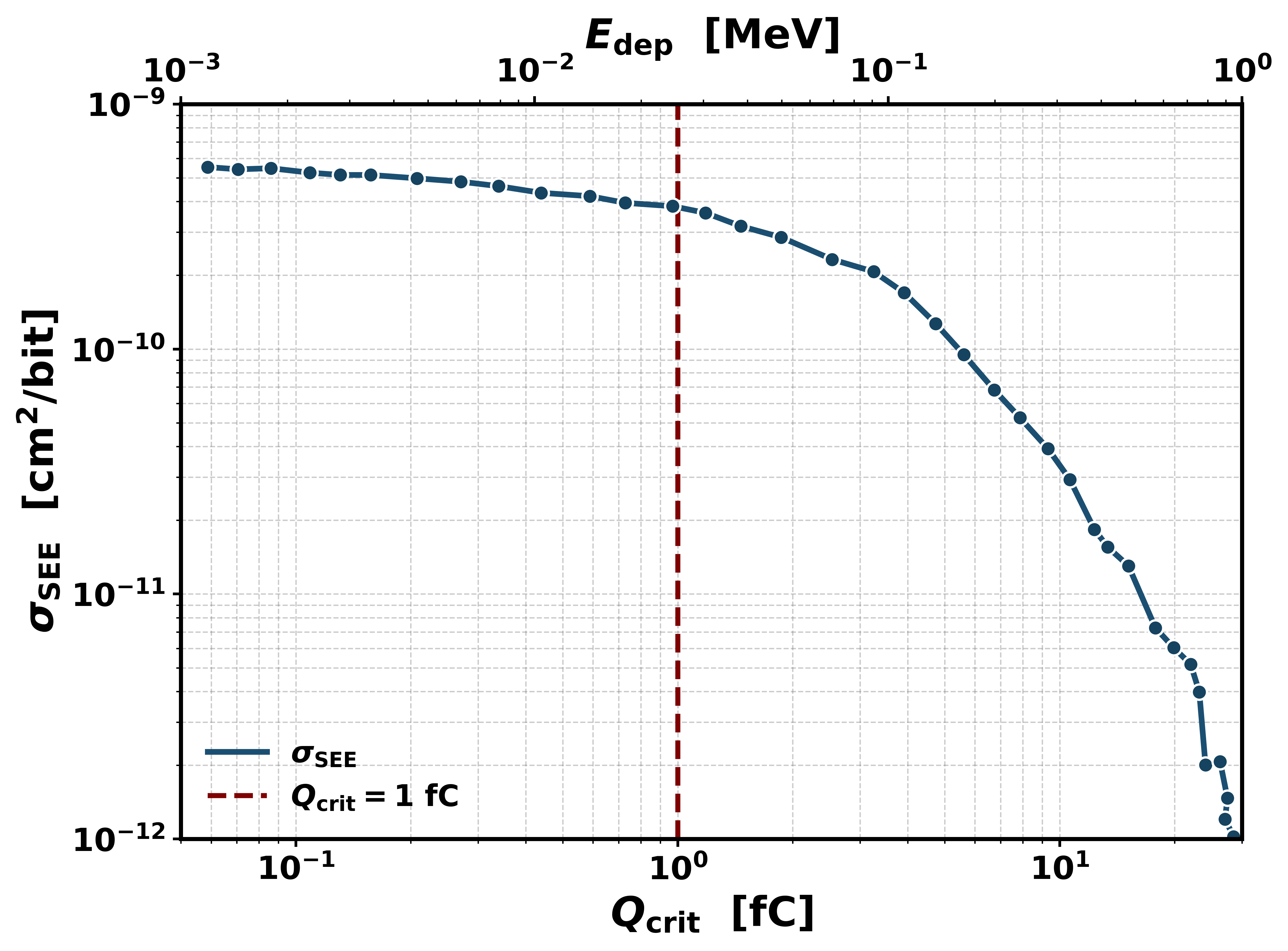}
\caption{Setup for FPGA device testing (left). Simulated SEE cross section in silicon vs. critical charge $Q_{\mathrm{crit}}$ at 10\,cm from source; vertical line marks $Q_{\mathrm{crit}}=1$\,fC (right).}
\label{fig:see}
\end{figure}

For electronics testing, we used FLUKA to compute the neutron spectrum at 10\,cm and fed it to G4SEE~\cite{lucsanyi2022} to simulate single event effect (SEE) cross sections in a silicon diode as a function of critical charge $Q_{\mathrm{crit}}$ (Fig.~\ref{fig:see}). The result is flat at $\sim10^{-9}$\,cm$^{2}$/bit for $Q_{\mathrm{crit}}<1$\,fC and falls off above, consistent with 14\,MeV data on submicron CMOS. We also irradiated commercial silicon devices and FPGA circuits at the CUP. Single bit-flips were observed in program memory, triggering automatic reconfiguration. Error rates scaled with firmware resource use: the FFT design ($\sim$86\% chip load) showed roughly an order of magnitude more errors than the compact LFSR design at the same fluence. These early results show that useful SEE data can already be collected at current yield levels.

\section{Summary and outlook}
\label{sec:summary}

We present NESSA as an operational, calibrated facility for fast neutron research. The GENIE\,16 generator has been commissioned and calibrated at $(4.33\pm0.29)\times10^{8}$\,n/s. The neutron field is reproduced consistently by four independent Monte Carlo codes, and $^{235}$U fission chamber response agrees with predictions within 10\% at two positions spanning two orders of magnitude in flux. Activation of structural materials and air is compatible with intended operation. Initial indium foil experiments and electronics exposures have been performed; the facility is now ready for nuclear data, detector response, and electronics studies. Upgrades include a pneumatic rabbit system, a thermalized beamline, and a higher-yield generator to $10^{10}$\,n/s.

\section*{Acknowledgment}
We thank the Swedish Energy Agency and the Swedish Center for Nuclear Technology (SKC) for financial support, and SKB for the GENIE\,16 loan. The contributions of the Carl Trygger Foundation are also acknowledged. We thank our colleagues from the Applied Nuclear Physics division and FREIA Laboratory, Uppsala University, for their support during commissioning.


\begin{thebibliography}{99}

\bibitem{leung2004}
K.N. Leung \textit{et al.}, Compact Neutron Generator Development and Applications (2004).

\bibitem{drosg2000}
M. Drosg, DROSG-2000: Neutron Source Reactions. IAEA report IAEA-NDS-87 Rev.~9 (2005).
 
\bibitem{sjostrand2014}
H. Sj\"ostrand \textit{et al.}, Total Monte Carlo Evaluation for Dose Calculations (2014).
  
\bibitem{elias2024}
E. Arnqvist, Neutron Spectrometry Using Activation Detectors. Uppsala University Student Thesis (2024).
 
\bibitem{nndc}
National Nuclear Data Center, Brookhaven National Laboratory, NuDat 3.0, \texttt{https://www.nndc.bnl.gov/nudat3/}
 
\bibitem{endf}
D.A. Brown \textit{et al.}, ENDF/B-VIII.0: The 8th major release of the nuclear reaction data library. Nucl. Data Sheets \textbf{148}, 1--142 (2018).
 
\bibitem{fluka2022}
C. Ahdida \textit{et al.}, New Capabilities of the FLUKA Multi-Purpose Code. Front. Phys. \textbf{9}, 788253 (2022).
 
\bibitem{agostinelli2003geant4}
S. Agostinelli \textit{et al.}, Geant4---a simulation toolkit. Nucl. Instrum. Meth. A \textbf{506}, 250--303 (2003).
 
\bibitem{shultis2013mcnp}
J.K. Shultis and R.E. Faw, \textit{An MCNP Primer} (2013).
 
\bibitem{niita2006phits}
K. Niita \textit{et al.}, PHITS---a particle and heavy ion transport code system. Radiat. Meas. \textbf{41}, 1080--1090 (2006).
 
\bibitem{ctreport}
G. Ericsson \textit{et al.}, Development of a neutron field monitoring system for the DT neutron generator facility NESSA. Report to the Carl Trygger Foundation, CTS 22:2218 (Uppsala University, 2026).
 
\bibitem{sunitha2020}
A.M. Sunitha \textit{et al.}, Measurements of $^{115}$In(n,2n)$^{114m}$In reaction cross sections at 14--14.54\,MeV. J. Radioanal. Nucl. Chem. \textbf{326}, 637--645 (2020).
 
\bibitem{luo2017}
J. Luo, L. Jiang and S. Li, Activation cross section and isomeric cross-section ratio for the (n,2n) reaction on $^{113,115}$In. Nucl. Sci. Eng. \textbf{188}, 198--206 (2017).
 
\bibitem{lucsanyi2022}
D. Lucs\'anyi \textit{et al.}, G4SEE: A Geant4-based single event effect simulation toolkit and its validation. IEEE Trans. Nucl. Sci. \textbf{69}, 273--281 (2022).
\end{thebibliography}
\end{document}